\documentclass[prl,a4paper,graphicx,preprint,reprint,twocolumn]{revtex4-1}

\usepackage{graphicx}
\usepackage{dcolumn}
\usepackage{bbm}
\usepackage{amsmath}
\usepackage{float}
\usepackage{lipsum}
\usepackage[colorlinks=true,
			allcolors=blue,bookmarks=false,pdfusetitle]{hyperref}
\usepackage{url}

\begin{document}

\title{Higher Order Quantum Ghost Imaging with Ultra-Cold Atoms} 

\author{S.~S.~Hodgman}

\email{sean.hodgman@anu.edu.au}

\author{W.~Bu}

\author{S.~B.~Mann}

\author{R.~I.~Khakimov}

\author{A.~G.~Truscott}

\affiliation{\normalsize{Research School of Physics and Engineering, Australian National University, Canberra 0200, Australia}}

\date{\today}
\begin{abstract}
Ghost imaging is a quantum optics technique that uses correlations between two beams to reconstruct an image in one beam from photons that do not interact with the object being imaged.  While pairwise (second order) correlations are usually used to create the image, higher order correlations can be utilized to improve the performance of ghost imaging.  In this paper, we demonstrate higher order atomic ghost imaging, using entangled ultracold metastable helium atoms from an s-wave collision halo.  We construct higher order ghost images up to 5th order and show that using higher order correlations can improve the visibility of the images without impacting the resolution.  This is the first demonstration of higher order ghost imaging with massive particles and the first higher order ghost imaging protocol of any type using a quantum source. 
\end{abstract}
\maketitle

Ghost imaging is an unconventional imaging method from quantum optics \cite{Erkmen2010,Shapiro2012}, which uses  
two correlated beams of photons.
One beam interacts with the object, after which the arrival time (only) of each photon from the beam is detected on a `bucket' detector, while the second photon from each pair is detected with full 3D spatial and temporal resolution on a multi-pixel detector, but never interacts with the object.  Using the correlations between the two beams, the image of the object can be reconstructed.  While the technique was first theoretically proposed \cite{Klyshko1988JETP,Belinskii1994} and experimentally demonstrated \cite{Pittman1995,Strekalov1995} using light, it has also recently been extended to X-rays \cite{Pelliccia2016}, cold atoms \cite{Khakimov2016} and electrons \cite{Li2018}, along with a recent proposal involving neutrons \cite{Chen2018}.  The correlated nature of ghost imaging means that in certain circumstances, such as for weakly absorbing objects \cite{Brida2010} or at low light levels \cite{Morris2015}, it can out-perform conventional imaging.  Ghost imaging has applications in a number of areas including optical encryption \cite{Kong2013,Yuan2016}, improved telecommunications \cite{Ryczkowski2016} and remote sensing \cite{Erkmen2012,Hardy2013}.  It also has the potential to reduce the dosage rates in imaging \cite{Pelliccia2016} and tomography \cite{Kingston2018} using radiation such as X-rays, where potential damage to the sample from the radiation is a concern.

There are two distinct types of ghost imaging that have been demonstrated.  Thermal or classical ghost imaging uses thermal or pseudothermal light, relying on semi-classical Hanbury Brown-Twiss (HBT) correlations to produce the images.  In contrast, quantum ghost imaging uses a quantum source of correlated pairs, which for photons is usually created via spontaneous parametric down conversion (SPDC).  While the SPDC pairs are entangled, entanglement is not necessary for ghost imaging \cite{Bennink2002,Erkmen2008}, although the performance of the imaging by some measures may be improved with entanglement \cite{Bennink2004}.

While the majority of ghost imaging implementations utilise pairwise (second order) correlations, the basic schemes can be extended to employ higher order correlations between more particles by incorporating additional detectors into the setup.  This has been shown to be able to improve the quality of key imaging parameters such as the visibility and contrast-to-noise ratio (CNR) \cite{Chen2010,Li2012}, as well as the resolution \cite{Zhou2012}, when implemented for ghost imaging with thermal light.    
However, due to the lower probability of higher order correlated events being detected, in practical cases such improvements do not necessarily translate to performance gains over second order ghost imaging when additional imaging improvements such as background subtraction are taken into account \cite{Chan2010}.  The use of higher order correlations to enhance the performance of ghost imaging is similar to the increase in visibility of multi-photon interference from thermal light via higher order correlations \cite{Agafonov2008}.  However, despite the body of work on higher order thermal ghost imaging, to the best of our knowledge there has been no demonstration of ghost imaging for quantum light.  This is partly because the relatively small two photon bunching amplitude for thermal ghost imaging limits the achievable visibility to $\frac{1}{3}$, meaning that there is a significant improvement attainable using higher order ghost imaging.  For quantum ghost imaging using SPDC sources, the low occurrence of high order correlations would make higher order ghost imaging challenging.

Here we demonstrate higher order ghost imaging using correlated pairs of ultra cold metastable helium (He*) atoms \cite{Vassen2012} from an s-wave scattering halo of two colliding Bose-Einstein condensates \cite{Perrin2007, Jaskula2010}. As shown in our previous experiment \cite{Khakimov2016}, by imposing a mask on one of the atoms in each pair and detecting those atoms with full 3D resolution, we construct ghost images using the time correlations between those atoms and their corresponding correlated partners that do not pass through the mask.  In addition to the two-atom correlations due to pairwise scattering, the complex many-body correlations in the halo leads to a complex hierarchy of correlations \cite{Hodgman2017}, and we use a range of these higher order correlations to construct higher order ghost images up to 5th order.  The quality of the resulting images was characterised via the visibility and resolution, with higher order imaging shown to be able to produce images with improved visibility at no detriment to resolution. 

The experimental procedure for our higher order ghost imaging is similar to our previous method for 2nd order ghost imaging \cite{Khakimov2016}, with the starting point being a Bose-Einstein Condensate (BEC) of $\approx 10^6$ helium atoms in the $m_J=+1$ sublevel of the long lived $2^3S_1$ metastable state \cite{Hodgman2009}, confined in a magnetic trap.  A Raman pulse then transfers nearly all atoms into the untrapped $m_J=0$ state, while also imparting a momentum of $\mathbf{K}=-\sqrt{2} k_0 \mathbf{\hat{z}}$ in the downwards direction (with gravity), where $k_0 = 2\pi/\lambda$ and $\lambda=1083.2$ nm is the wavelength of the Raman laser, and allowed to fall under gravity.  The untrapped BEC is then split into 12 momentum components, also in the $\mathbf{\hat{z}}$ direction, by a second diffraction pulse operating in the Kapitza-Dirac regime \cite{Khakimov2016}.  Each pair of BECs in adjacent momentum components then collide, producing a spherical halo with a radius (in the momentum space frame of reference centred on the two BEC components) of $k_{r}\approx k_0/\sqrt{2}$ comprising pairs of back-to-back correlated \cite{Perrin2007,Hodgman2017} and entangled \cite{Shin2018} atoms, analogous to the pairs of photons produced from a SPDC source in quantum optics.  Each of the 11 scattering halos has a radial gaussian width of $w\approx 0.03k_r$ with an average mode occupancy varying from $n = 0.002(2)$ to $n = 0.08(1)$ across the halos, depending on the relative fraction of BEC atoms transferred to each of the momentum modes. The multiple halos technique is used to reduce our data acquisition time, with the ghost imaging implemented for each halo separately.  Within each halo there is a complex hierarchy of higher order correlations, with correlations due to both the scattering collision and HBT style multi-particle interference (see \cite{Hodgman2017} for details).  In general, the degree of correlation increases for higher orders of correlations.  

\begin{figure}[hbt]

	\includegraphics[width=0.45\textwidth]{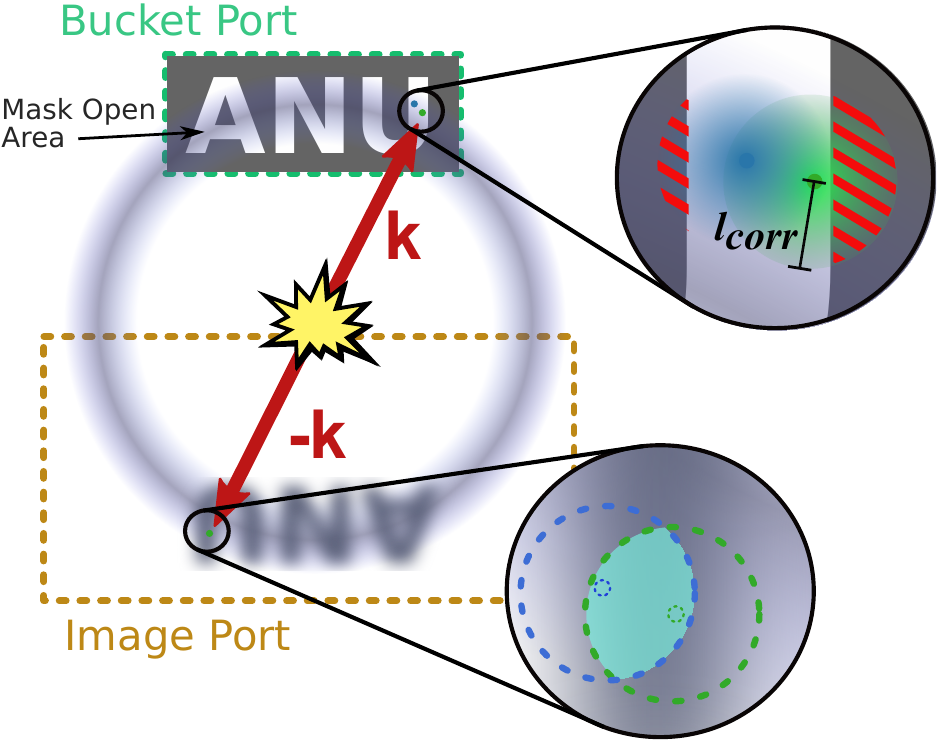}
	\caption{Schematic of atomic higher order ghost imaging (in this case third order).  A halo (grey shaded ring) of back-to-back correlated atoms is produced via pair-wise $s$-wave collisions to produce correlated atoms between momenta $\mathbf{k}$ and $\mathbf{-k}$.  The detector is divided in two, with atoms on the `bucket port' side of the halo passed through a mask (here the letters `ANU'), with only the arrival times of atoms that pass through recorded and all spatial information of these atoms discarded.  These arrival times are then correlated with the arrival times of atoms in the `image port', which are measured with full 3D resolution, to match groups of 2, 3, 4 or 5 atoms around $\mathbf{k}$ and $\mathbf{-k}$, depending on the order of ghost imaging being used.  The $x$ and $y$ components of these image port atoms form the ghost image.  Insets show the improvement in visibility possible with higher order imaging schemes, where additional correlated atoms detected in the bucket port constrain the location of the atom in the image port (in this case the cyan area). This makes the image atoms more likely to be found in the corresponding image location, rather than outside it, where they would increase the background (see main text for details).} 
	\label{fig:schematic}
\end{figure}

\begin{figure*}
	\includegraphics[width=0.85\textwidth]{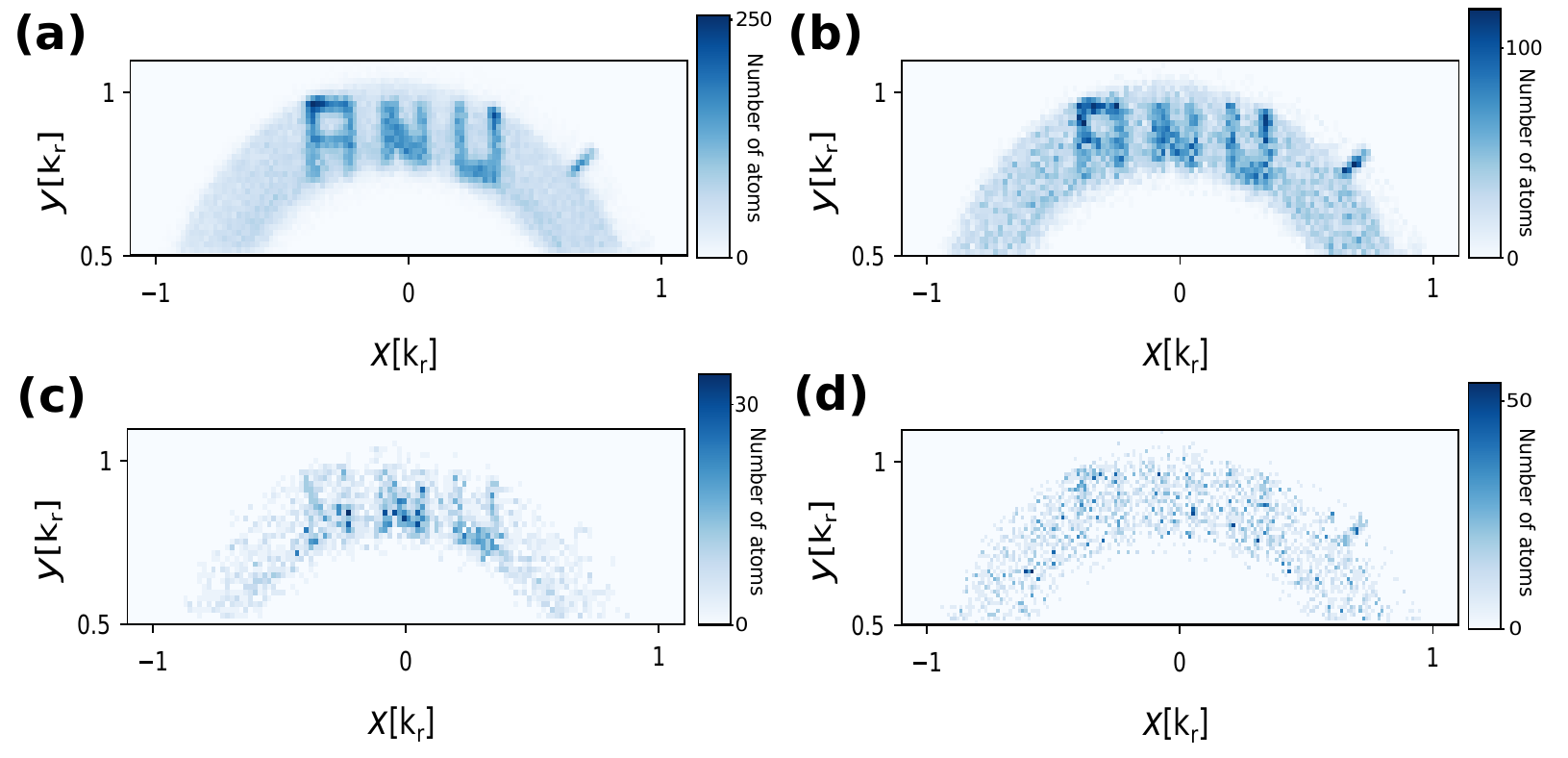}

	\caption{\label{fig:comp} A selection of atomic ghost images formed in the detector image port using a transmission mask consisting of the letters ``ANU" to cover the detector bucket port.  There is only one case (a) for second order ghost imaging, consisting of 1 atom in each port.  In contrast, for three atom ghost imaging there are two configurations, with either two atoms in the image port (b) or two atoms in the bucket port (c).  Higher orders have correspondingly more possible configurations, of which one is shown (d) for four atoms, with three of these atoms detected in the image port.  The background semi-circular arc corresponding to the halo is visible in each image and is due to random, uncorrelated atoms.  The scale of the images is given in terms of the collisional recoil momentum $k_r$, while the number of atoms in each pixel is shown by the colour bars for each image.  }
	\label{fig:ANU-120bins}
\end{figure*}

The halos fall under gravity $\approx\!850$\!~mm onto a multi-channel plate and delay-line detector, while also expanding during the fall time.  This detector allows the position of the atoms at the detector, corresponding to the atomic momenta in the collision halo, to be measured in full 3D, with an $x,y$ resolution of $\sim\!~\!120$~$\mu$m and a resolution in the $z$ direction of $\sim 12\mu$m ($\equiv 3\mu$s in arrival time) \cite{Henson2018}.   The detector is divided in half centred on the halo, with one half assigned as the `bucket' port and the other the `multi-pixel' port (see Fig. \ref{fig:schematic}).  A software mask \cite{mask_note} is imposed on atoms arriving in the bucket port, so that only atoms that arrive in a defined region (the open area of the mask) are recorded.   All the spatial information of these bucket port atoms is then discarded, with only their arrival times retained.  In the basic scheme of second order ghost imaging, for each of these bucket port atoms (which have momentum $\mathbf{k}=k_x\hat{\mathbf{x}}+k_y\hat{\mathbf{y}}+k_z\hat{\mathbf{z}}$), any atom in the multi-pixel port which arrives within an arrival time interval set by the correlation length of the scattered pairs ($l_{corr}\approx 0.03k_r$ \cite{Hodgman2017}) centred around -$k_z\hat{\mathbf{z}}$ is then added to the ghost image.  The image is created from all correlated ghost image atoms across all 11 halos from $\sim$45,000 different experimental runs.  Fig. \ref{fig:ANU-120bins}(a) shows a sample image of a mask of the letters `ANU'.  

For $N$th order ghost imaging the procedure is similar, with $N$ total atoms required to be detected distributed in some particular combination around $ k_z\hat{\mathbf{z}}+\mathbf{\Delta z}$ and $-k_z\hat{\mathbf{z}}+\mathbf{\Delta z}$,  with all $|\mathbf{\Delta z} | < l_{corr}$.  If this condition is met then all atoms around $-k_z\hat{\mathbf{z}}$ are added to the image.  Each $N$ has different possible combinations with varying degrees of correlation \cite{Hodgman2017}.  For example, in third order ghost imaging one atom is always in each of the mask and the image, with the choice for the extra atom to be added to either port.  Higher order extensions follow the same pattern.  The two third order cases are shown in Fig. \ref{fig:ANU-120bins}(b) and (c), while the fourth order case of three atoms in the image and one in the mask port shown in Fig. \ref{fig:ANU-120bins}(d).  

To quantitatively analyse and compare the performance of the different orders and combinations of ghost imaging, two widely used measures were calculated for each case: the visibility and the resolution.  A rectangle mask of dimensions (0.22$\times$0.18)$k_r$ was used for both of these, to simplify the analysis, although the results are similar for other mask shapes.

The visibility $V$ of the ghost image is defined as
\begin{equation}
    V= \frac {I-B} {I+B}.
\end{equation}
$I$ is the total number of atoms in the region of the ghost image corresponding to the location of open area of the mask on the opposite side of the halo, while $B$ is the total number of atoms in the region of the semi-circular arc corresponding to the halo in the rest of ghost image.  The non-halo regions, where almost no atoms will be found, are excluded.  Both $I$ and $B$ are normalized individually by the size of their respective areas.

\begin{figure}[hbt]
	\includegraphics[width=0.45\textwidth]{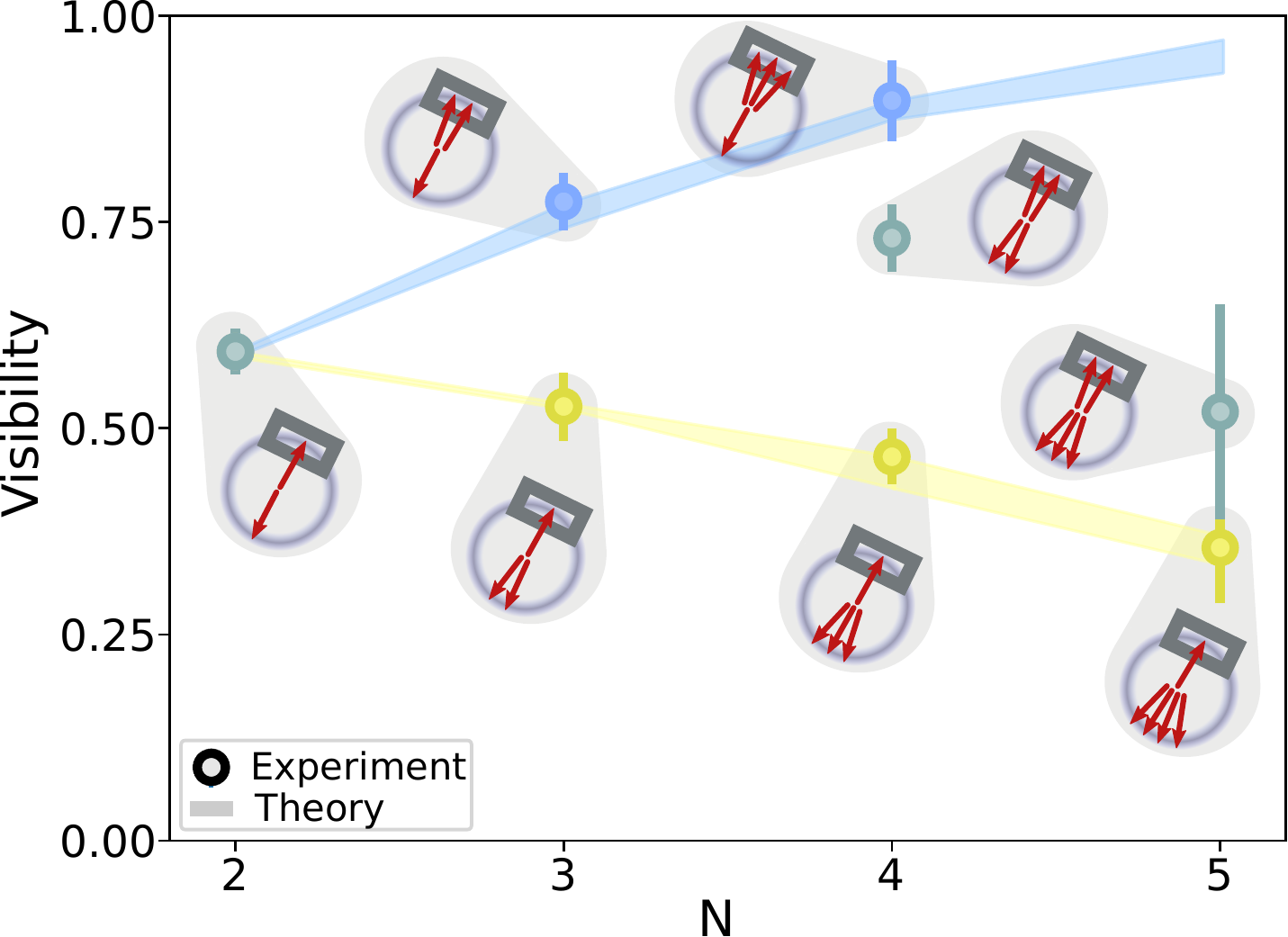}
	\caption{\label{fig:comp}  The experimentally measured (circles) and numerically simulated (bands) visibilities for a range of different types and orders ($N$) of ghost imaging, as shown in the schematic insets.  Blue visibilities are for the optimal cases, with one atom in the image port and the rest in the bucket (mask) port, with the reverse cases shown in yellow (green points are for intermediate cases). Error bars are obtained by combining the variance in intensity $I$ and background $B$.
	The variance of $B$ and $I$ are caused by the variance in the total number of hits in the halo in different experiments.  
	} 
\label{fig:vis_plot}
\end{figure}

The experimentally measured visibility for 8 different ghost imaging cases is shown in Fig. \ref{fig:vis_plot}, representing all combinations up to 5th order (the two cases of fifth order with 1 and 2 atoms in the image port are not shown, as there were insufficient counts to produce a meaningful result). As the plot shows, the best visibility for each order $N$ is achieved for the cases with one atom in the image port and the rest in the mask (blue points).  For these cases $V$ increases with $N$, while for the opposite case (one atom in mask port and the rest in the image port) shown in yellow, $V$ decreases with $N$.  For the intermediate cases (shown in green) the results are somewhere in the middle.  The insets in Fig. \ref{fig:schematic} show schematically the process (for the three atom case) of how higher order ghost imaging can improve the visibility.  In this case, the higher degree of correlation for three atoms means that a second atom (blue) detected in the bucket port is more likely to be found within a correlation length $l_{corr}$ of the first (green).  Intuitively, this increased detection probability results from HBT style bunching of the two particles in the bucket port.  Due to the mask, these will both by definition lie within the image, as potential arrival locations outside the mask (indicated by the red hatched area in Fig. \ref{fig:schematic}) are excluded as a possible location for additional atoms.  An atom detected in the image port, however, is more likely to be found within a radius around -$\mathbf{k}$ of $\sim l_{corr}$ on the opposite side of the halo (cyan area).  The two atoms detected confine this to a narrower window with (on average) a larger overlap on the image than the area allowed around $l_{corr}$ for a single bucket port atom only (green dashed line).

In both of the extreme cases (blue and yellow in Fig. \ref{fig:vis_plot}), the degree of correlation is the same, although $N-1$ times as many atoms are detected in the image port for the case with one atom in the image port.  However, for higher order schemes some intermediate (green) configurations are more correlated than others (e.g. for fourth order the degree of correlation scales as $24+24/n+4/n^2$ in the case of two atoms on each side of the halo but as $24+18/n$ for the cases of three atoms on each side \cite{Hodgman2017}, which is smaller for the values of $n$ in our halos).  This would suggest that the change in $V$ is not primarily due to an increased degree of correlation, but rather the relative number of atoms in the image port compared to the bucket port.  The cause of this is that in the best case more atoms in the mask increases the probability that the only atom in the image will be a correlated hit and not a background atom, as the atoms in the mask are limited to a smaller area.  In contrast, there is a much larger area where atoms in the ghost image can possibly fall, and thus a lot more background atoms are included.  The trend of the extreme cases is similar to that observed with optics for multi-photon interference  \cite{Agafonov2008}.  

This interpretation of extra atoms in the image causing increased blurring due to finite $l_{corr}$, and this effect being more dominant than the relative bunching amplitude, is supported by numerical simulations of our ghost imaging \cite{SOMs}.   The visibilities generated for our experimental parameters are shown as yellow and blue bands in Fig. \ref{fig:vis_plot}, which agree well with the experimental results.  Importantly, the simulations show that while an increased bunching amplitude $g^{(N)}_{BB}$ does improve visibility, above some value $g^{(N)}_{crit}$ this effect mostly saturates, whereas increasing $l_{corr}$ always decreases the visibility.  For our experiments we are always in the regime $g^{(N)}_{BB}\gg g^{(N)}_{crit} $ \cite{SOMs}, and thus the optimal visibility is always with the maximum number of atoms in the bucket port.

The other image quality measure implemented was the resolution, which is dependent on the finite correlation length of the correlated atoms \cite{Khakimov2016}. To measure the resolution the ghost image of a $(0.22\times0.18)k_r$ size rectangle for each case is integrated along the $y$ axis.  A Gaussian convolved with a top-hat function of the size of the square mask was then fitted to the data and the width of the fitted Gaussian $\sigma$, which corresponds to the resolution of the image, extracted. 

\begin{figure}[bt]

	\includegraphics[width=0.45\textwidth]{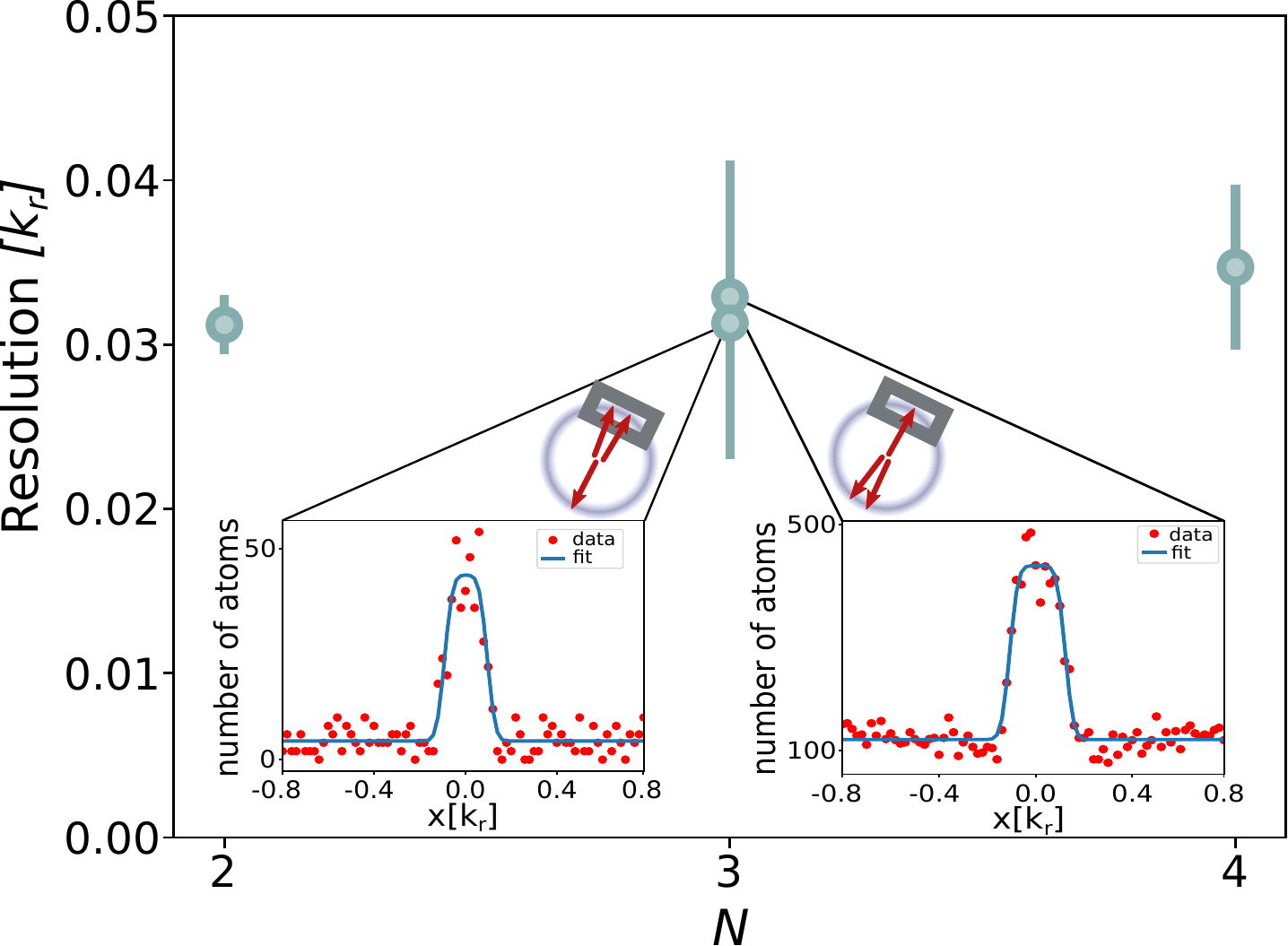}
		
	\caption{Resolution of ghost imaging using different orders $N$. The resolution is calculated by forming a ghost image of a square mask, integrating along the $y$ dimension and then fitting a Gaussian function convolved with a top hat of the width of the imaged square. Insets show these plots for the two third order cases (the left inset shows the case of one atom in the image, the right inset is two atoms in the image).  In all cases the resolution does not significantly change for different methods within error.}
	\label{fig:res_summary}
\end{figure}

The measured resolutions for the different cases are shown in Fig. \ref{fig:res_summary}, along with some sample plots showing the raw data.  As can be seen, the measured resolution for all cases is close to $l_{corr}$.  There is also little difference in the measured resolution for the different ghost imaging methods, although the resolution for higher order cases with more atoms in the image may be slightly worse.

In conclusion, we have demonstrated higher order ghost imaging with atoms up to 5th order, the first such demonstration with massive particles and the first higher order ghost imaging experiment using a quantum source, as to the best of our knowledge all previous demonstrations were for semi-classical ghost imaging implemented with thermal or pseudo-thermal light.  The visibility was seen to improve with higher orders for the cases with only 1 atom in the ghost image, while the reverse occurred for cases with only one atom in the mask.  In all cases the resolution did not change significantly. 
The improved visibility makes higher order ghost imaging better for applications with demanding visibility or signal to flux ratio requirements, such as in X-ray ghost imaging \cite{Pelliccia2016} and tomography \cite{Kingston2018}, where higher-order schemes could be used to produce images with improved visibility at lower dosage rates, or similar damage sensitive schemes with atoms such as atomic lithography.  By extending the higher order atomic ghost imaging scheme presented here, complex fundamental tests of quantum mechanics with massive particles could also be possible, such as multi-atom entanglement~\cite{Kofler2012} or Bell's inequality measurement schemes\cite{Jack2009} using 3 or more particles.

\begin{acknowledgments}
 The authors would like to thank Bryce Henson and David Shin for technical assistance. This work was supported through Australian Research Council (ARC) Discovery Project grants DP120101390, DP140101763 and DP160102337. SSH is supported by ARC Discovery Early Career Researcher Award DE150100315.
\end{acknowledgments}

\bibliography{Refs}

\pagebreak

\section{Supplementary Information}

\subsection*{Numerical Simulation}

To further investigate the experimentally observed trends in the ghost imaging visibility and gain additional insights into the factors contributing to this, we performed a simple Monte Carlo simulation that aims to emulate the key features of ghost imaging.  
The simulations used the same rectangle mask (0.22$\times$0.18)$k_r$ used to produce the experimental visibility plot shown in Fig. \ref{fig:vis_plot}.  For two atom ghost imaging, a pair of random 2 dimensional coordinates were generated within the range of the mask, representing the atom recorded in the bucket port. 
In the neighborhood centred on the inverse of each of these coordinates (i.e. the opposite side of the halo), a pair of random coordinates was generated following a 2D Gaussian distribution with width $l_{corr}$, representing the correlated atoms detected in the image port.  
To account for uncorrelated atoms, a proportion of image port atoms were chosen to be randomly distributed over the entire image port, with the exact fraction corresponding to the degree of correlation ($g^{(2)}$) in the experimental data.  Once sufficient random co-ordinates were generated, the visibility was calculated in the same manner as for the experimental data. 20 datasets were randomly generated for each parameter set, to produce an uncertainty spread of the range of visibilities for each case.  The parameters varied were 
the correlation length ($l_{corr}$) and ratio of correlated atoms to atoms in the background (i.e. the degree of correlation).

\begin{figure}[H]
    \centering
    \includegraphics[width=0.45\textwidth]{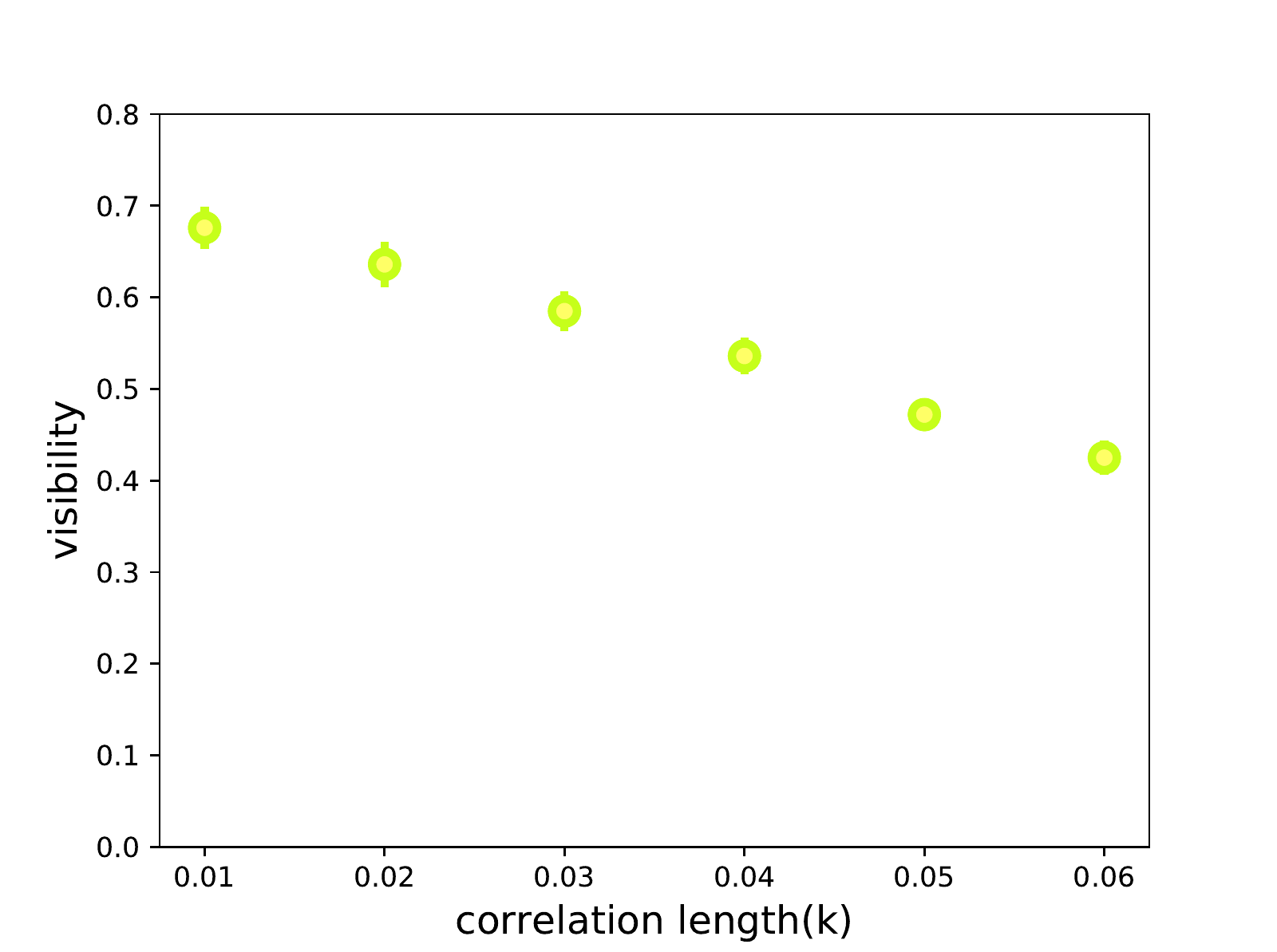}
    \caption{\label{dz}Monte-Carlo simulation results for two atom ghost imaging showing the scaling of the visibility of the ghost image with different correlation lengths in units of the halo radius $k_r$.}   
    
\end{figure}
Fig. \ref{dz} shows the results of the first parameter investigated, $l_{corr}$ using the rectangle mask with 2 correlated atoms. As the correlation length increases, naturally more atoms in correlated pairs would fall outside of the corresponding image area, which would increase the background atom numbers resulting in a decrease of visibility of the image.  Note that the experimental halo has $l_{corr}\sim0.03k_r$.

\begin{figure}[H]
    \centering
    \includegraphics[width=0.45\textwidth]{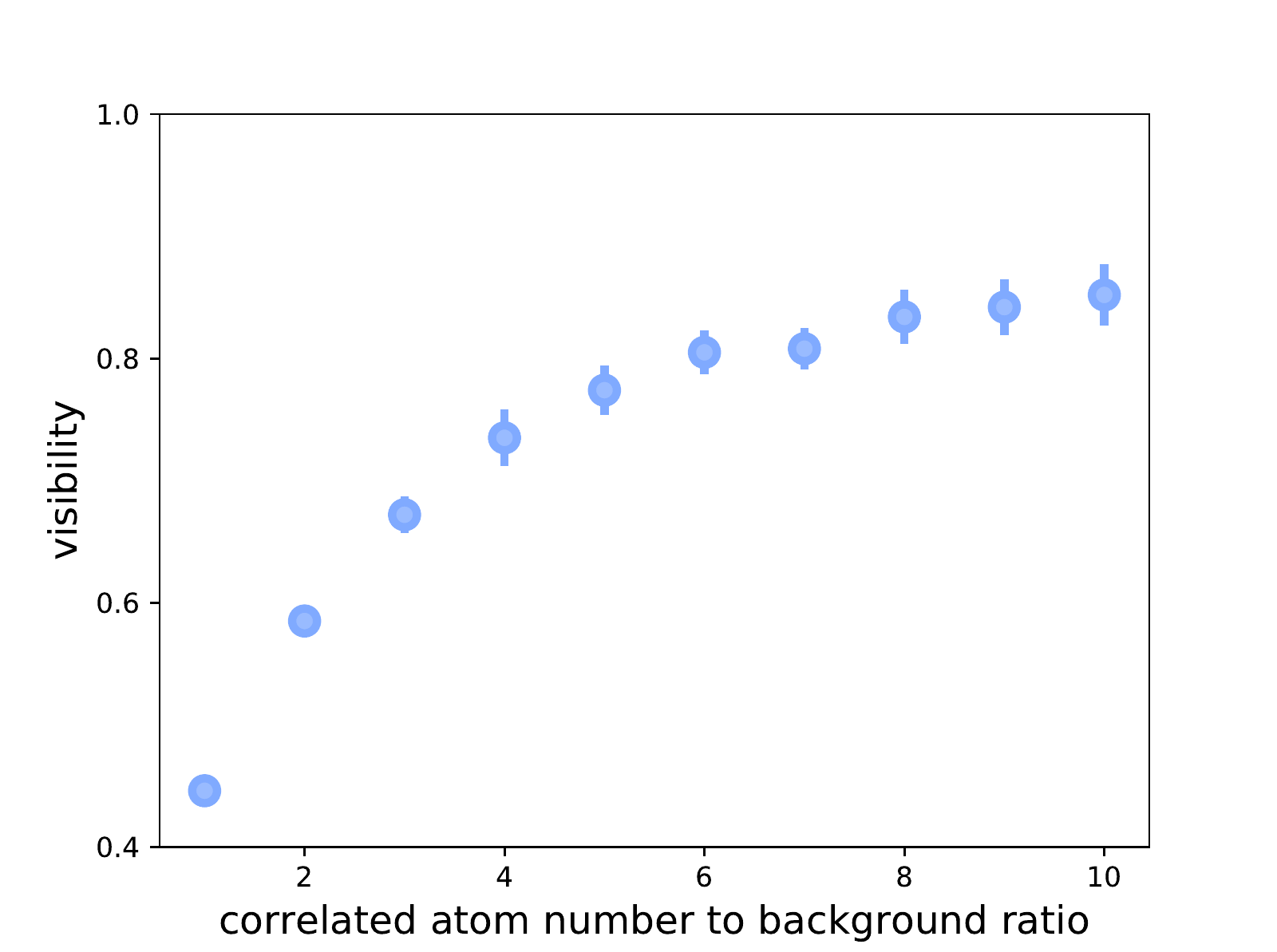}
    \caption{\label{background}The plot shows the visibility of the ghost image with two atom ghost imaging from Monte-Carlo simulations using a rectangle mask (see text for details).  The data is generated with different degree of correlated atoms, here represented by the ratio of correlated atoms to background atoms in the image ($\propto g^{(2)}$).}

\end{figure}

Setting $l_{corr}=0.03k_r$ and varying the number of atoms in the background, Fig. \ref{background} shows the dependence of visibility on the degree of correlation, represented theoretically by the ratio of correlated image atoms to uniformly distributed background atoms.  The degree of correlation corresponds to the experimental observable of the second order correlation function $g^{(2)}$, 
which is directly related to the ratio between correlated atom number and background atom number.
This conversion is for every $m$ correlated pairs if one atom is added to the background then $g^{(2)} = m +1$, since in the simulations the background and image areas are same size.
While Fig. \ref{background} exhibits a trend of increasing visibility as the degree of correlation increases, it asymptotically approaches a limit for high $g^{(2)}$ values (for this parameter set around $v=0.85$).  This limit is due to factors such as the values of the correlation length, the ratio of uncorrelated to correlated atoms, the size of the mask etc. 
Since these factors will somewhat depend on the experiment, to compare with the experimental data in Fig. \ref{fig:vis_plot} the degree of correlation is adjusted as a free parameter in the two atom data to ensure the generated visibility matches the experimental value, then kept consistent for the higher order data.

With $l_{corr}$ set to the experimental value and the degree of correlation adjusted to match the experimental 2-atom result, the various different permutations of higher order ghost imaging were simulated.
For these higher order cases, additional atoms were added to the image and/or bucket ports for each correlated event as required, with the previously described gaussian distributions with width $l_{corr}$ centered around the original bucket port atom or the opposite side of the halo.  The results for the two limiting cases of one atom in the bucket port (yellow line) and one atom in the image (blue line) are shown in Fig. \ref{fig:vis_plot}, with the shaded region indicating the spread of numerical values generated.  As can be seen the results match the experimental data very well for all orders, suggesting that our simple simulation captures the key aspects of the process.

\end{document}